\documentclass[%
onecolum,
 aip,
 amsmath,amssymb,
reprint,
]{revtex4-1}
\usepackage[ansinew]{inputenc}
\usepackage{graphicx}
\usepackage{dcolumn}
\usepackage{bm}
\usepackage{amssymb,amsmath,latexsym,amsfonts}
\usepackage{amsthm,subfigure}
\usepackage{amsbsy}
\usepackage[mathscr]{eucal}
\usepackage[dvipdf]{epsfig}
\usepackage{color}

\begin{document}

\title{Drops in the wind: their dispersion and  COVID-19 implications}
\author{Mario Sandoval}
\email{sem@xanum.uam.mx}
\author{Omar Vergara}
\affiliation{
 Department of Physics, Universidad Autonoma Metropolitana-Iztapalapa, 
 Mexico City 09340, Mexico.
}

\begin{abstract}
Most of the works on the dispersion of droplets and their COVID-19
(Coronavirus disease) implications address droplets' dynamics in quiescent
environments. As most droplets in a common situation are immersed in
external flows (such as ambient flows), we consider the effect of canonical  flow profiles namely, shear flow, Poiseuille flow, and unsteady shear flow
on the transport of spherical droplets of radius ranging from 5$\mu$m to 100 
$\mu $m, which are characteristic lengths in human talking, coughing or sneezing
processes. The dynamics we employ satisfies the Maxey-Riley (M-R) equation. An 
order-of-magnitude estimate allows us to solve the M-R equation to leading order
analytically, and to higher order (accounting for the Boussinesq-Basset
memory term) numerically. Discarding evaporation, our results to leading order indicate that the
maximum travelled distance for small droplets ($5\mu m$ radius) under a shear/Poiseuille
external flow with a maximum flow speed of $1m/s$ may easily reach more than 250
meters, since those droplets remain in the air for around 600 seconds. The maximum
travelled distance was also calculated to leading and higher orders, and it is observed
that there is a small difference between the leading and higher order
results, and that it depends on the strength of the flow. For example, this difference
for droplets of radius $5\mu m$ in a shear flow, and with a maximum wind speed of $5m/s$, is
seen to be around $2m$. In general, higher order terms  are observed to
slightly enhance droplets' dispersion and their flying time.  
\end{abstract}

\date{\today }
\maketitle

\section{INTRODUCTION}

So far, COVID pandemic is still growing, as of 18:34pm Central European Time
(CET), February $6$ $2021$, there have been $104,956,439$ corroborated cases
of COVID-19, encompassing $2,290,488$ deaths, reported to the World Health
Organization (WHO). It is believed that airborne transmission is one of the
main mechanisms for COVID spreading\cite{molina,6} and that potential sources of infected droplets are breathing, sneezing, coughing or simply talking. Unfortunately, most of the reported
literature neglects the effect of ambient flows on droplets transport. These
flows are often present in daily activities such a wind flows, ventilation
generated currents in offices, homes, malls, among other public places.

Recent studies suggest that talking may be among the most dangerous
mechanisms for generating infected droplets\cite{molina,11,AbkarianPNAS,Katy}%
. According to Tan \cite{Katy} and Bourouiba\cite{Bourouiba}, speech induced
plumes can travel $1.3m$ in $2s$ or even $8m$, which is a distance way
longer than the $2m$ social distancing. Abkarian and Stone \cite{11} using
high-speed imaging showed that pronouncing consonants (typical to many
languages) such as 'Pa','Ba', and 'Ka' are potent aerolization mechanisms.
Abkarian \textit{et al}\cite{AbkarianPNAS} using theory, experiments, and
simulations documented the flow generated after speaking and breathing,
which is in fact the responsible for droplets' transport. 
Notice that the 2m social distancing, only considered quiescent flows\cite%
{Jones,AbkarianPNAS,Bourouiba}. This situation is not always satisfied in
daily conditions, where wind is frequently present either naturally or
induced by air condition in buildings or even by simple motion of people\cite%
{linden}. Further evidence of airborne transmission of COVID-19 disease possible enhanced by air currents  are: A reported infection of $96$ people out of $216$
working in a eleventh floor in a call center of South Korea\cite{park}; a
singing rehearsal in Washington, where $53$ singers were infected even they
were located in a volleyball court but ventilation air currents were present%
\cite{Katy}. A study precisely on the effectiveness of air ventilation and
COVID implications\cite{linden} suggests that only certain type of
ventilation called 'displacement ventilation' properly designed to extract
contaminated hot air, could be the most effective air condition mechanism to
reduce the risk of infection. 
Related works where droplets produced by breathing, sneezing, coughing, or
talking, and immersed in external flows are few. Some examples are Cummins 
\textit{et al} \cite{1} who studied micrometric spherical droplets in the
presence of a source-sink flow, which simulated a scenario where droplets
are produced and subjected to an extraction mechanism (air condition).
Incorporating evaporation, humidity, and an uniform external wind, Dbouk and
Drikakis\cite{2} presented a conjugated heat and flow transfer problem
(occurring in a saliva droplet) and coupled to a CFD analysis. They proposed a
transient Nusselt number and varied relative humidity (RH), temperature, and
speed of flow. Contrary to past studies, they conclude that evaporation of
droplets is enhanced at low RH and high temperatures. As an example, they
results indicate that in a cloud of droplets in an environment at RH=$50\%$,
temperature $T<30^oC$, and under an external flow of $4km/h$, there will not be
evaporation. Their simulation however, only reached five seconds, hence the
full dispersion of droplets could not be reached. B. Blocken \textit{et al}%
\cite{28} also performed a CFD analysis for people exhaling while walking or
running, and emanating from them possible infected droplets. However, their
simulation only considered droplets of radius $20\mu m$ and beyond. As it
has recently been observed, the smallest the size of a droplet, the more
dangerous it may be\cite{26}. Feng \textit{et al}\cite{30} located two virtual humans
$3.05m$ apart and let one human to eject droplets while coughing or sneezing.
Using a computational particle fluid dynamics model that considered
evaporation, external wind and condensation, and even Brownian motion, they
found that at this distance, potentially infected droplets easily reach hair
and face of the other human. They also performed a study on the filtration
efficiency of several masks. In conclusion, most of the previous works
suggest that the safe distancing is not enough under wind conditions.

In this work, we consider micrometrical spherical droplets emanating from a
person while talking either normally (exit speed of droplets around $1m/s$) or
strongly (exit speed of droplets around $5m/s$)\cite{AbkarianPNAS,1}, and
under the presence of external flow currents, and calculate, following the
Maxey-Riley (M-R) equation \cite{7}, their effect on the maximum spreading of
droplets of constant radius ranging from $5\mu m$ to $100\mu m$. The wind
currents profiles are modeled as a shear flow, a Poiseuille flow, and an
unsteady shear flow that considers the typical time-dependent situation of
wind blowing and ceasing. 
The present study is organized as follows: Section \ref{PM} describes the
model, and the order-of-magnitude of each term in the M-R equation. Section %
\ref{no} presents analytical results for the dispersion of micrometric
droplets subject to three external flow profiles namely, shear, Poiseuille,
and unsteady shear flow. Here, the Boussinesq-Basset memory force is neglected. Section %
\ref{yes} is intended to study, by performing numerical simulations, the
effect of the Boussinesq-Basset memory force (and the same external flows as
in Sec. \ref{no}) on the dispersion of droplets. Discussions and conclusions are offered in
Sec. \ref{conclus}.

\begin{figure}
\includegraphics [width=5cm]{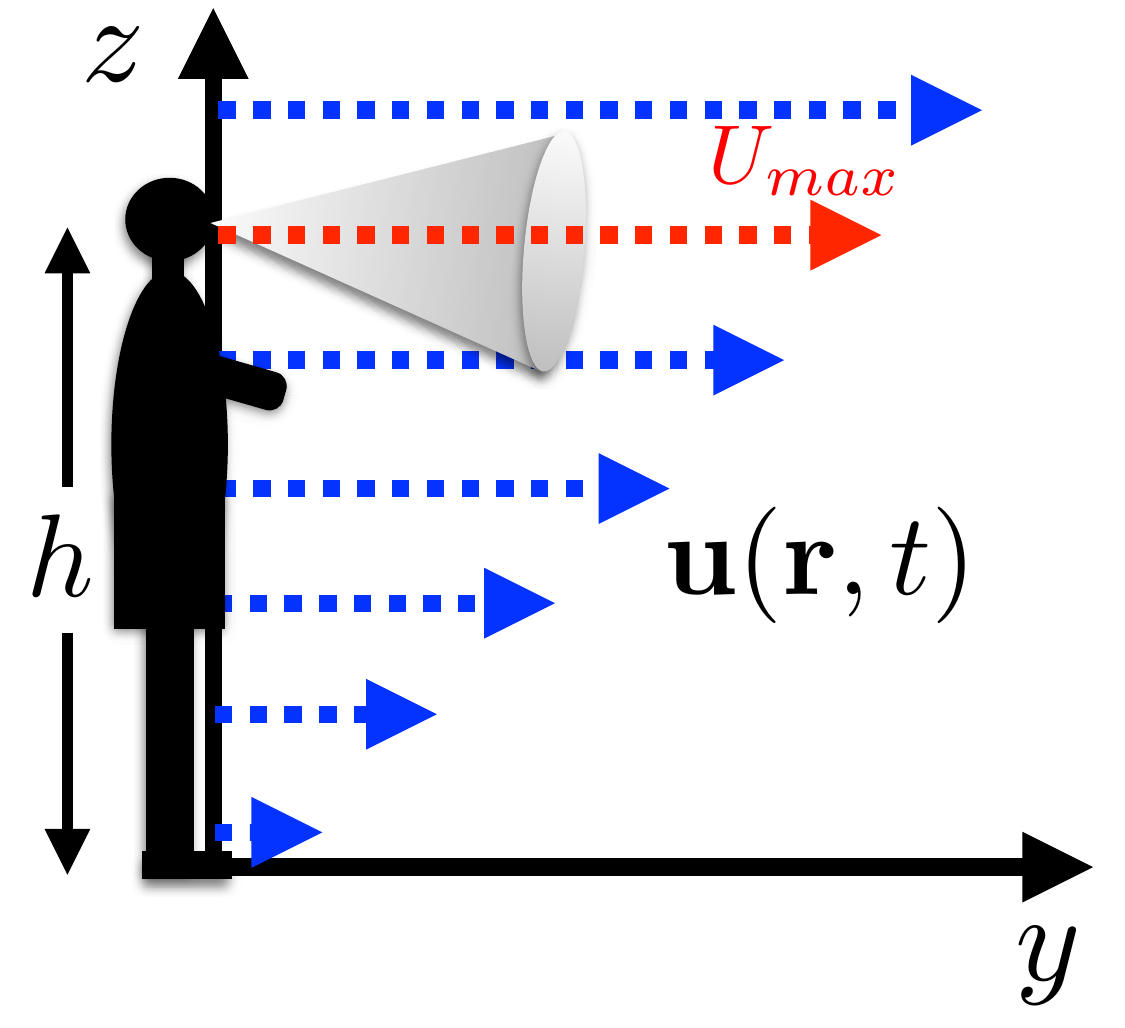}
\caption{Schematic of the problem studied. A person talking either normally
(exit speed of droplets around $1m/s$) or strongly (exit speed of droplets
around $5m/s$)\protect\cite{AbkarianPNAS,1}, and under the presence of an
external flow \textbf{u}(\textbf{r},t).}
\label{model}
\end{figure}

\begin{figure*}
\includegraphics [width=16cm]{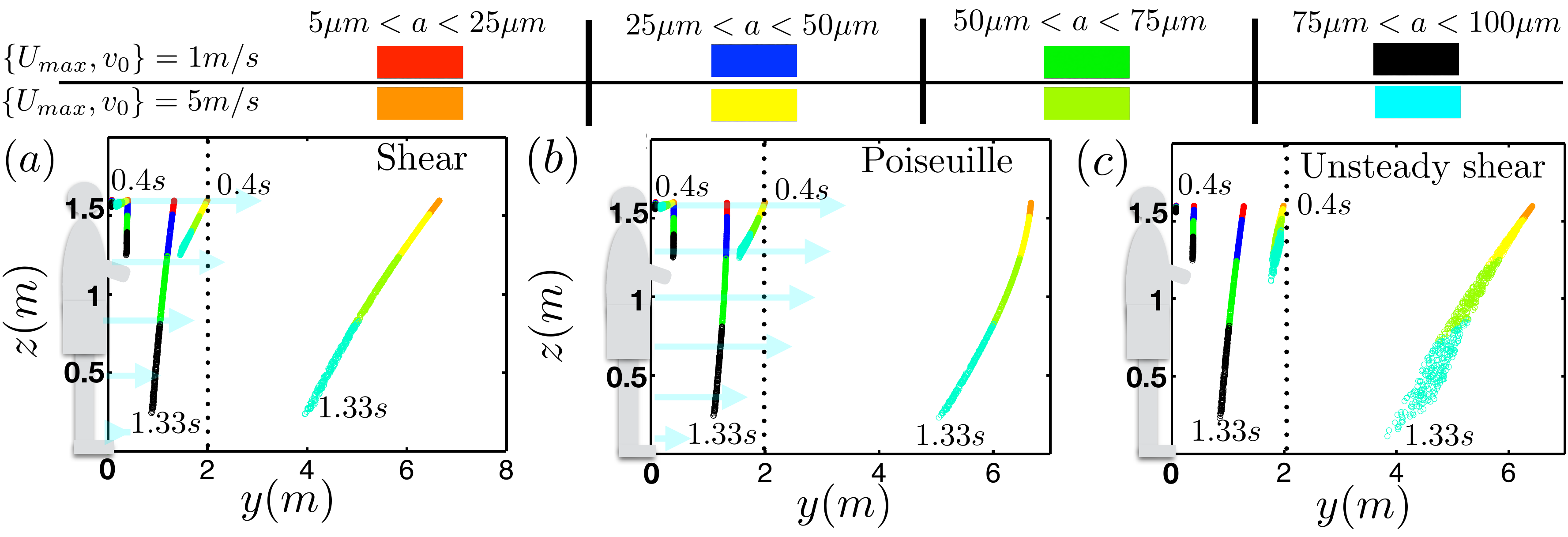}
\caption{ A person talking and spreading micrometric droplets of radius $a$
ranging from 5 $\protect\mu m$ to 100 $\protect\mu m$. The spreading is
under the presence of an external flow \textbf{u}, and it is shown for three
different times after the droplets were ejected from a cone shape of velocities modeling a person's mouth
namely, $t=0.08s,t=0.4s$, and $t=1.33s$. (a), Drops' distribution during a
normal and strong talk with exit speed of $v_{0}=1m/s$ and $v_{0}=5m/s$,
respectively; and under a shear flow with $U_{max}=\{1m/s, 5m/s\}$ (see
color code). (b) The same as in (a) but under the presence of a Poiseuille
flow. (c) The same as in (a) but under the presence of an unsteady shear flow.
The safe distancing is indicated as a vertical dotted-black line. The light
blue arrows represent the flow profile.}
\label{model}
\end{figure*}

\section{Physical model}

\label{PM}

Let us analyze the motion of noninteracting spherical droplets of mass $m$,
radius $a$, in an environment with air density $\rho _{a}$ and air viscosity 
$\mu _{a}$, under gravity $\mathbf{g}$, and subject to an external flow of
the form $\mathbf{u}\left( \mathbf{r},t\right) $, with a characteristic
speed $U_{max}$, where $\mathbf{r}(t)=(x,y,z)$ is the droplet's position,
here $t$ represents time. In this study we will be considering three flows
namely shear, Pouiseuille, and unsteady shear flow. Droplets immersed in
these profiles will have a characteristic speed $v(t)$ which decreases as
the droplets fall due to gravity. With the latter physical quantities a
Reynolds number ($\mathfrak{Re}$) can be defined as $\mathfrak{Re}=\rho
_{a}(U_{max}-v)a/\mu _{a}$ which satisfies $\mathfrak{Re}<1$. Therefore,
these droplets individual dynamics described by their translational
velocity, $\mathbf{v}(t)=(v_{x},v_{y},v_{z})$, follows the Maxey-Riley
equation\cite{7} 
\begin{eqnarray}
m\frac{d\mathbf{v}}{dt} &=&\left( m-m_{a}\right) \mathbf{g+}m_{a}\left. 
\frac{D\mathbf{u}}{Dt}\right\vert _{\mathbf{r}}  \notag \\
&&-\frac{m_{a}}{2}\frac{d}{dt}\left( \mathbf{v}_{T}+\frac{a^{2}}{15}\left.
\nabla ^{2}\mathbf{u}\right\vert _{\mathbf{r}}\right)   \notag \\
&&-R_{T}\mathbf{v}_{T}-R_{T}a\frac{\mathbf{v}_{T}(0)}{\sqrt{\pi \upsilon
_{a}t}}  \notag \\
&&-R_{T}a\int_{0}^{t}\frac{d\mathbf{v}_{T}}{d\tau }\frac{d\tau }{\sqrt{\pi
\upsilon _{a}\left( t-\tau \right) }},  \label{MR}
\end{eqnarray}%
where $m_{a}$ is the mass of air displaced by the sphere, $\upsilon
_{a}=\mu _{a}/\rho _{a}$ represents the kinematic viscosity, $R_{T}=6\pi
a\mu _{a}$ is the resistance coefficient; while $\rho $ indicates the
droplet's density. In addition, the following definitions are included: 
\begin{eqnarray}
\mathbf{v}_{T} &=&\mathbf{v-}\left. \mathbf{u}\right\vert _{\mathbf{r}}-%
\frac{a^{2}}{6}\left. \nabla ^{2}\mathbf{u}\right\vert _{\mathbf{r}}
\label{sn} \\
\frac{D\mathbf{u}}{Dt} &=&\frac{\partial \mathbf{u}}{\partial t}+\mathbf{%
u\cdot \nabla u,}  \label{d1} \\
\frac{d\mathbf{u}}{dt} &=&\frac{\partial \mathbf{u}}{\partial t}+\mathbf{%
v\cdot \nabla u.}  \label{d2}
\end{eqnarray}%
The forces on the right hand side of \ Eq. (\ref{MR}) are respectively, the droplet's
weight; bouyancy force; forces in the undisturbed flow due to local pressure
gradients;  the added or virtual mass force; Stokes drag force; and the last
two terms represent the Boussinesq-Basset memory force. To find the
order-of-magnitude of each term in the Maxey-Riley Eq. (\ref{MR}), one can
introduce the following dimensionless quantities $\overline{\mathbf{r}}=%
\mathbf{r}/a,\overline{\mathbf{v}}=\mathbf{v/}U,\overline{t}=\left(
U/a\right) t,$ where the characteristic speed is defined as $U=mg/R_{T},$
which is the terminal velocity of a falling droplet. After applying those
dimensionless variables to Eq. (\ref{MR}), one arrives at 
\begin{eqnarray}
Re\frac{d\overline{\mathbf{v}}}{d\overline{t}} &=&-\frac{9\lambda }{2}\left(
1-\lambda \right) \mathbf{k+}\lambda Re\left. \frac{D\overline{\mathbf{u}}}{D%
\overline{t}}\right\vert _{\overline{\mathbf{r}}}  \notag \\
&&-\frac{\lambda Re}{2}\frac{d}{d\overline{t}}\left( \overline{\mathbf{v}}%
_{T}\mathbf{+}\frac{1}{15}\left. \overline{\nabla }^{2}\overline{\mathbf{u}}%
\right\vert _{\overline{\mathbf{r}}}\right)   \notag \\
&&-\frac{9\lambda }{2}\overline{\mathbf{v}}_{T}-\frac{9\lambda \sqrt{Re}}{2%
\sqrt{\pi }}\frac{\overline{\mathbf{v}}_{T}(0)}{\sqrt{\overline{t}}}  \notag
\\
&&-\frac{9\lambda \sqrt{Re}}{2\sqrt{\pi }}\int_{0}^{\overline{t}}\frac{d%
\overline{\mathbf{v}}_{T}}{d\overline{\tau }}\frac{d\overline{\tau }}{\sqrt{%
\left( \overline{t}-\overline{\tau }\right) }}.  \label{gen}
\end{eqnarray}%
where the Reynolds number $%
Re=\rho _{a}Ua/\mu _{a}$ has been introduced, as well as the ratio $\lambda
=m_{a}/m=\rho _{a}/\rho =1.225\times 10^{-3}$. Here $\overline{\mathbf{g}}=\mathbf{g}a/U^{2}$. 

\section{Dynamics without Boussinesq-Basset memory}

\label{no}

Let us start analyzing the order of magnitude estimate of the terms in Eq. (%
\ref{MR}) based on typical droplets ejected after talking. According to
recent studies\cite{AbkarianPNAS,1,3}, the radius of those droplets range
between $5\mu m$ to $100\mu m$, hence Table \ref{T1} shows their respective
Reynolds number and the order-of-magnitude involved in Eq. (\ref{gen}). 
\begin{table}
\caption{Order-of-magnitude for different terms appearing in the M-R Eq. (%
\protect\ref{gen}) as a function of the radius $a$. }
\label{T1}%
\begin{ruledtabular}
\begin{tabular}{ccccd}
$a(\mu m)$&$Re$&{ $\lambda \sqrt{Re}$}&{$\lambda Re$}\\
\hline

5 &$ 8.4\times 10^{-4} $&$ 3.5\times 10^{-5} $& $1\times 10^{-6}$ \\ 
10 & 0.0067 &$ 1\times 10^{-4} $& $8.2\times 10^{-6}$ \\ 
20 & 0.05 &$ 2.83\times 10^{-4} $& $6.55\times 10^{-5} $\\ 
30 & 0.18 &$ 5.2\times 10^{-4} $& $2.2\times 10^{-4} $\\ 
50 & 0.83 & 0.0011 & 0.001 \\ 
100 & 6.7 & 0.0032 & 0.0082\\ 
\end{tabular}
\end{ruledtabular}
\end{table}
From this table and  for droplets of radius $a\in \left[ 5\mu m,20\mu m%
\right]$, the limit $%
Re\rightarrow 0$ can be applied to the M-R Eq. (\ref{gen}) to finally get 
\begin{equation}
\overline{\mathbf{v}}_{T}=-\mathbf{k},
\end{equation}%
which \ implies that the dynamics of droplets of this size is 
\begin{equation}
\overline{v}_{x}=0;\overline{v}_{y}=\overline{u}_{y},\overline{v}_{z}=-1.
\end{equation}%
The latter result belongs to the so-called overdamped approximation, where
inertia does not play a role, and particles immediately reach the external
flow speed. Alternatively, by keeping inertia and neglecting terms of order $%
\sqrt{Re}$ and higher, the Maxey-Riley equation can be rewritten as%
\begin{equation}
\frac{d\overline{\mathbf{v}}}{d\overline{t}}=-\overline{g}\left( \mathbf{k}+%
\overline{\mathbf{v}}_{T}\right) .  \label{SE}
\end{equation}%
This equation allows us to observe the dynamics of droplets at short times.
\subsection{ Shear and Poiseuille flows as external wind}
Let us solve Eq. (\ref{SE}) under the presence of shear and Poiseuille flows
modeling the effect of wind on the spherical droplets. These profiles are
given by%
\begin{equation}
\mathbf{u}=\left( U_{\max }\frac{z}{h}+U_{0}z\left( z-h\right) \right) \;%
\mathbf{j.}  \label{prof}
\end{equation}%
Notice that the shear flow profile correspond to the case $U_{0}=0.$ Given Eq. (\ref%
{prof}), we solve Eq. (\ref{SE}) whose solution including dimensions and
subject to $\mathbf{r}(0)=(0,0,h)$ and $\mathbf{v}(0)=(v_{x0},v_{y0},v_{z0})$
can be shown to be
\begin{eqnarray}
v_{x}(t) &=&v_{x0}e^{-Kt},\text{ \ }v_{z}(t)=\gamma _{z}e^{-Kt}-\frac{g}{K},
\label{vx} \\
v_{y}(t) &=&\gamma _{y}e^{-Kt}-U_{0}\left( \frac{\gamma _{z}}{K}\right)
^{2}e^{-2Kt}+U_{0}\frac{\gamma _{z}g}{K}t^{2}e^{-Kt}  \notag \\
&&-Ate^{-Kt}+B-Ct+U_{0}\frac{g^{2}}{K^{2}}t^{2},  \label{vy}
\end{eqnarray}
\begin{figure*}
\includegraphics [width=17cm]{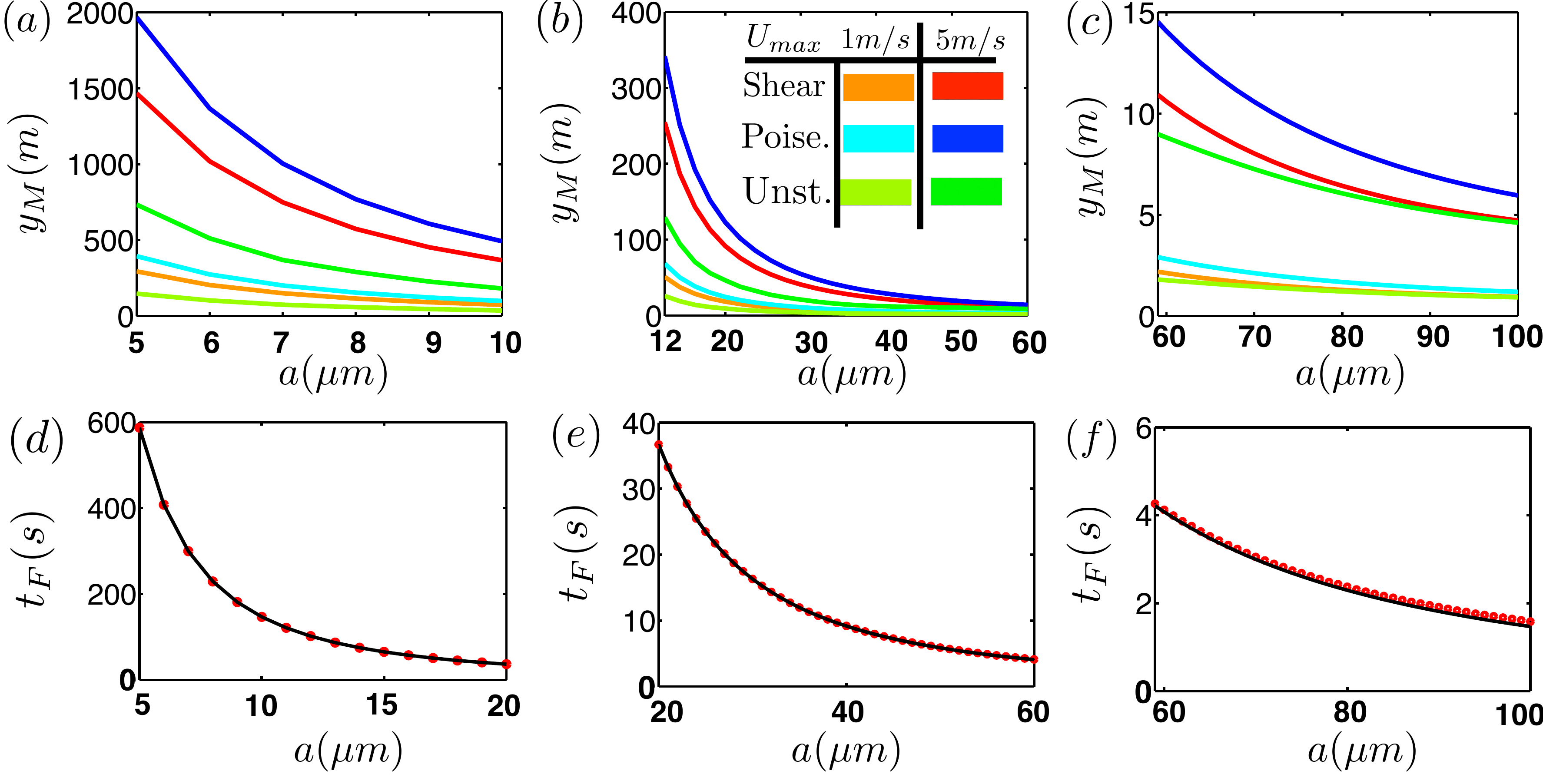}
\caption{(a)-(c) Maximum dispersion ($y_{M}$) along the $y-$%
direction of spherical droplets as a function of the radius $a$, as well
as a function of different external flows namely, shear, Poiseuille, and
unsteady shear flow. The color code in (b) is valid for all figures. (a)
Maximum dispersion for $a$ ranging from 5$\protect\mu m$ to 10$\protect\mu m$%
. (b) Maximum dispersion for $a$ ranging from 12$\protect\mu m$ to 60$\protect%
\mu m$. (c) Maximum dispersion for $a$ ranging from 60$\protect\mu m$ to 100$%
\protect\mu m$. (d)-(f) Flying time ($t_{F}$) of spherical droplets
as a function of the radius $a$. Here, $v_z(0)=0$. Red circles represent the
whole $z-$component in Eq. (\protect\ref{xz}), while the black-solid line
represents the approximated equation $t_F\approx {9\protect\mu_a h}/{2%
\protect\rho ga^2}$. (d) $a$ ranging from 5$\protect\mu m$ to 20$\protect\mu m
$. (f) $a$ ranging from 20$\protect\mu m$ to 60$\protect\mu m$. (e) $a$
ranging from 60$\protect\mu m$ to 100$\protect\mu m$.}
\label{distance}
\end{figure*}
where $K=R_{T}/m$ and $\beta =K{a^{2}}U_{0}/3$. Integrating the velocities,
we get
\begin{eqnarray}  
x(t) &=&-\frac{v_{x0}}{K}e^{-Kt}+s_{x},\text{ \ }z(t)=-\frac{\gamma _{z}}{K}%
e^{-Kt}-\frac{g}{K}t+s_{z},   \notag \\
&& 
 \label{xz}
 \\
y(t) &=&\frac{1}{K}\left( \frac{A}{K}-\gamma _{y}-\frac{2U_{0}\gamma _{z}g}{%
K^{3}}\right) e^{-Kt}  \notag \\
&&+\frac{1}{K}\left( A-\frac{2U_{0}\gamma _{z}g}{K^{2}}\right) te^{-Kt}-%
\frac{U_{0}\gamma _{z}g}{K^{2}}t^{2}e^{-Kt}  \notag \\
&&+\frac{U_{0}\gamma _{z}^{2}}{2K^{3}}e^{-2Kt}+s_{y}+Bt-\frac{C}{2}t^{2}+%
\frac{U_{0}g^{2}}{3K^{2}}t^{3},  \label{y}
\end{eqnarray}%
where constants $A,B,C,\gamma _{y},\gamma _{z},s_x,s_y,s_z$ are defined in
Appendix \ref{A1}.

\subsection{ Unsteady shear flow as external wind}

A more realistic situation, is to consider the fact that air flows for a
while, and then stops, and then flows again. The simplest model is to assume
a time-dependent shear flow scenario. This unsteady vector flow field $%
\mathbf{u}(\mathbf{r},t)$ should satisfy the time-dependent Navier-Stokes
equations, which after assuming $\mathbf{u}(\mathbf{%
r},t)=(0,u_{y}\left( z,t\right) ,0)$, incompressibility, and a null pressure, reduce to
\begin{equation}
\frac{\partial u_{y}}{\partial t}=\upsilon _{a}\frac{\partial ^{2}u_{y}}{%
\partial z^{2}}\mathbf{.}  \label{p}
\end{equation}%
This parabolic equation must satisfy $u_{y}\left( z,0\right) =\left( U_{\max
}/h\right) z,u_{y}\left( 0,t\right) =0,$ and $u_{y}\left( h,t\right)
=U_{\max }/2\left( 1+\cos \omega t\right) $. Its solution can be veryfied to
be%
\begin{equation}
u_{y}\left( z,t\right) =\sum_{m=1}^{\infty }D_{m}g(t)\sin \frac{m\pi z}{h}+%
\frac{U_{\max }}{2h}\left( 1+\cos \omega t\right) z,  \label{solun}
\end{equation}%
where 
\begin{eqnarray}
g(t) &=&e^{-d_{m}t}+\frac{d_{m}}{\omega }\sin \omega t-\cos \omega t,
\label{un2} \\
D_{m} &=&\frac{\omega f_{m}}{d_{m}^{2}+\omega ^{2}},  \label{un3} \\
f_{m} &=&\left( -1\right) ^{m+1}\omega U_{\max }/m\pi ,  \label{un4} \\
d_{m} &=&\upsilon _{a}\left( \frac{m\pi }{h}\right) ^{2}.  \label{un5}
\end{eqnarray}%
Once Eq. (\ref{solun}) is available, it is plugged into Eq. (\ref{SE}) and
its solution subject to $\mathbf{r}(0)=(0,0,h)$ and $\mathbf{v}%
(0)=(v_{x0},v_{y0},v_{z0})$ although very lengthy, can be analytically
found. The dominant terms of the solution along the $y-$component for long
times read
\begin{figure*}
\includegraphics [width=16cm]{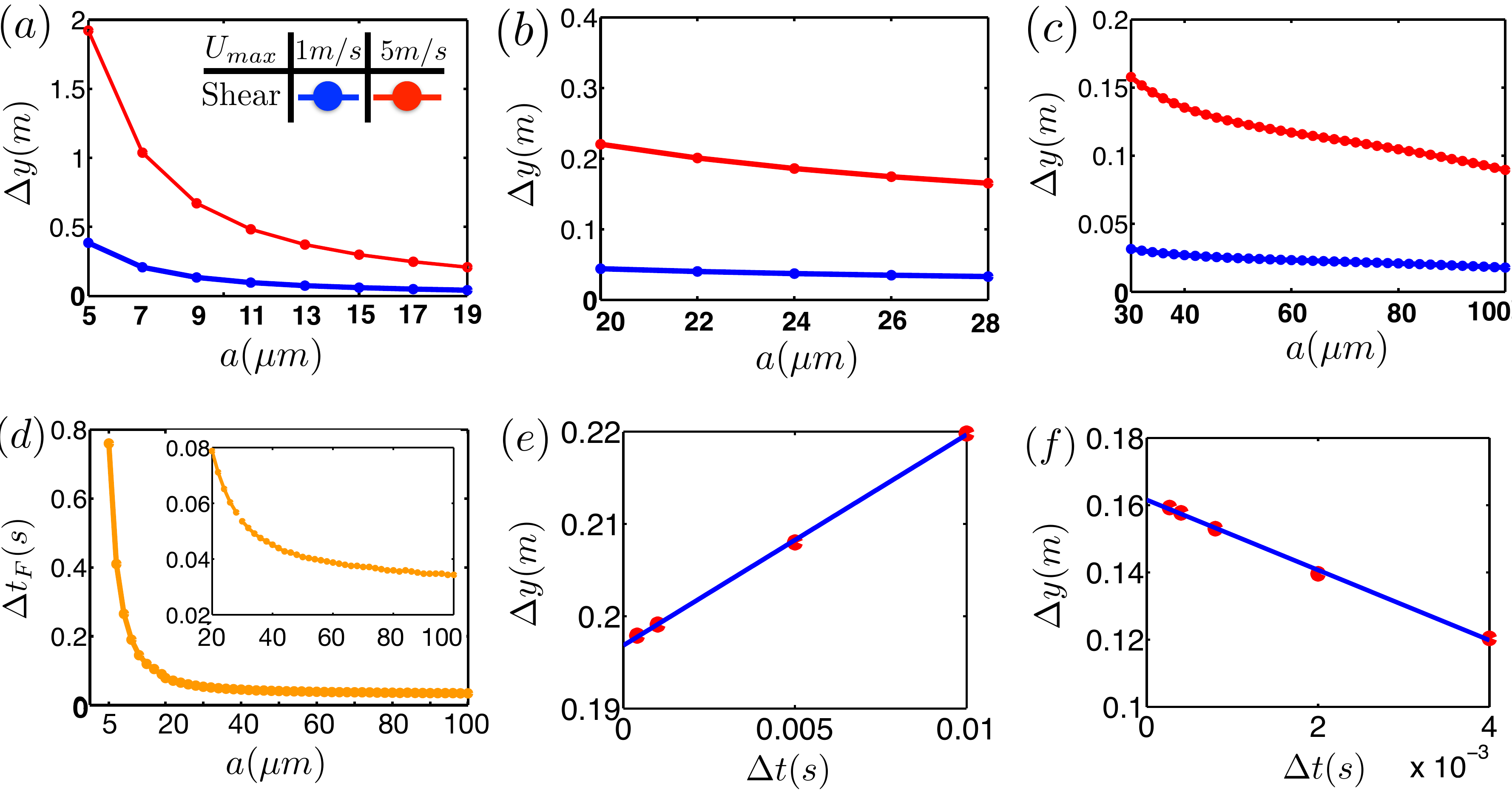} 
\caption{(a)-(c) Effect of higher order terms on the dispersion
along the $y-$direction of spherical droplets as a function of radius $a
$ and an external flow. Here $\Delta y=y_{MB}-y_{M}$, where $y_{MB}$ is the
maximum travelled distance by a droplet along the $y$-direction containing
first order terms ($a\leq 20\protect\mu m$) or the full terms in the M-R
equation ($a>20\protect\mu m$), whereas $y_{M}$ represents the solution
directly obtained from Eqs. (\protect\ref{xz}) and (\protect\ref{y}). (d)
Effect of higher order terms on the flying time of spherical droplets as a
function of radius $a$. Here $\Delta t_{F}=t_{FB}-t_{F}$, where $t_{FB}$
is the flying time by a droplet containing first order terms ($a\leq 20%
\protect\mu m$) or the full terms in the M-R equation ($a>20\protect\mu m$);
whereas $t_{F}$ represents the the solution directly obtained from Eqs. (%
\protect\ref{xz}). (e)-(f) Linear convergence of $\Delta y=y_{MB}-y_{M}$
as the time-step in the simulations decreases. (e) Convergence for a droplet
of radius $30\protect\mu m$ after discretizing the full M-R equation (%
\protect\ref{gen}). (f) Convergence for a droplet of radius $19\protect\mu m$
after discretizing the overdamped M-R equation (\protect\ref{overMR}).}
\label{basset}
\end{figure*}
\begin{eqnarray}
v_{y}(t) &=&-\frac{U_{\max }g}{2hK}t-\frac{U_{\max }g\left[ K\cos (\omega
t)+\omega \sin (\omega t)\right] }{2h(K^{2}+\omega ^{2})}t,  \notag \\
&&
\label{vun} 
\\
y(t) &=&\frac{U_{\max }}{2h}\left\{ \frac{g}{K^{2}}t-\frac{g}{2K}%
t^{2}\right.   \notag \\
&&\left. -\frac{g}{(K^{2}+\omega ^{2})}\left[ \frac{K}{\omega }t\sin (\omega
t)-t\cos (\omega t)\right] \right\} .  \label{posun}
\end{eqnarray}%
Once  Eq. (\ref{xz}), Eq. (\ref{y}), and Eq. (\ref{posun}) are available, we can now
exemplify a typical droplets' dispersion (cloud) after a person talks and generates
micrometric droplets (radius between 5 $\mu m$ to 100 $\mu m$ \cite%
{AbkarianPNAS,1}), which are subject to either a shear flow (Fig. \ref%
{distance}(a)), a Pouiseuille flow (Fig. \ref{distance}(b)), or an unsteady
shear flow (Fig. \ref{distance}(c)). We simulate the situation of a normal
and a strong talk by imposing an exit speed of droplets (from a cone shape of velocities representing a person's mouth)
of $v_{0}=1m/s$ and $v_{0}=5m/s$, respectively. The cloud is made of $1000$ droplets of random size  between 5 $\mu m$ to 100 $\mu m$. Additionally, we
impose a moderate ($U_{max}=5m/s$) and a calm ($U_{max}=1m/s$) wind
scenario. The cone shape as the initial velocity distribution, and as observed
from experiments\cite{1}, is implemented by imposing $\mathbf{v}%
(0)=(v_{0}\cos \varphi _{0}\sin \theta _{0},v_{0}\sin \varphi _{0}\sin
\theta _{0},v_{0}\cos \theta _{0})$, whose angular polar and azimuthal
initial extremum are set to $\theta _{0}^{max}=110^{o}$, $\theta
_{0}^{min}=81^{o}$, $\varphi _{0}^{max}=110^{o}$, and $\varphi
_{0}^{min}=81^{o}$. The case of a Poiseuille flow profile considers for a calm
flow $\{U_{max},U_{0}\}=\{1m/s,-0.4m^{-1}s^{-1}\}$ wheread for a moderate flow $%
\{U_{max},U_{0}\}=\{5m/s,-2m^{-1}s^{-1}\}$: The simulations for the
unsteady flow profile consider $\omega =0.5s^{-1}$. Finally, $(\varphi
,\theta )$ will randomly vary between their extremum. The results can be
visualized in Fig. \ref{distance} which shows the distribution of droplets
at three different times namely, $%
t=0.08s$, $t=0.4s$, and $t=1.33s$. A color code indicating the droplets' sizes
is also introduced. The safe distancing is indicated as a vertical
dotted-black line. Clearly, droplets move beyond the safe distancing. As it
can be seen, droplets under a moderate wind and after only $1.33s$, are
already $6m$ away from the person's mouth. Droplets under a calm wind
and less than $50\mu m$ in radius will overpass the safe distancing in the
next second. These results indicate a potential danger for people in a public space since the safe distancing has been easily surpassed. 
We finally take Eq. (\ref{xz}), Eq. (\ref{y}), and Eq. (\ref{posun}) to find
the maximum travelled distance ($y_{M}$) along the $y-$direction as a
function of droplets' size, as well as a function of different external
flows namely, shear, Poiseuille, and unsteady shear flow. The results are
shown in Fig. Figs. \ref{distance}(a)-(c). Figures \ref{distance}(a)-(c)
show that droplets under a Pouiseuille profile reached the the longest
distance compared to the other analyzed profiles. This figure also indicates
that for $U_{max}=5m/s$, droplets of radius $5\mu m$, and under a Poiseuille
flow, can travel until $2000m$, while around $1500m$ under a shear flow. As
expected, the condition of having a wind blowing and ceasing (unsteady flow)
reduces the droplets maximum dispersion to around $750m$ for $U_{max}=5m/s$.
On the other hand, our results to leading order indicate that the maximum
travelled distance for small droplets ($5\mu m$ radius) under a
shear/Poiseuille external flow with a maximum speed of $U_{max}=1m/s$ may
easily reach more than 250 meters. The long distance achieved by these small
droplets is because they remain in the air for around 600 seconds, see Figs. %
\ref{distance}(d)-(e). In these figures, the flying time obtained from Eq. (%
\ref{xz}), is shown as a solid-black line; while the red circles represent
the approximated equation $t_{F}\approx {9\mu _{a}h}/{2\rho ga^{2}}$. These
figures also indicate that the largest ($100\mu m$ radius) droplets can only
stay in the air for about $1.5s$. It is worth mentioning that all the
studied droplets ($a\in \left[ 5\mu m,100\mu m\right] $) reached a constant
vertical terminal velocity $U\approx mg/R_{T}$.

\section{Dynamics with Boussinesq-Basset memory}

\label{yes}

In the literature, the Boussinesq-Basset (B-B) memory force has been less
frequently considered, probably because of its order-of-magnitude and
because of the required computational effort. However, there exist some
theoretical\cite{12,15,7,20,22,23}, numerical\cite{14,17,5,21} and
experimental\cite{13,27} works dealing with this force. 
In this section we analyze the effect of the memory force term on the
dispersion an flying time of spherical droplets.

Consider first small droplets ranging between $a\in \left[ 5\mu m,20\mu m%
\right] $. According to Table \ref{T1}, their motion can be modeled by the
M-R equation under the overdamped approximation. However, by keeping terms of order $%
\sqrt{Re}$ to see the effect of the B-B memory force, Eq. (\ref{gen}) reads 
\begin{equation}
\overline{\mathbf{v}}_{T}=-\left( 1-\lambda \right) \mathbf{k}-\sqrt{\frac{Re%
}{\pi }}\int_{0}^{\overline{t}}\frac{d\overline{\mathbf{v}}_{T}}{d\overline{%
\tau }}\frac{d\overline{\tau }}{\sqrt{\left( \overline{t}-\overline{\tau }%
\right) }}.  \label{overMR}
\end{equation}%
To solve this integro-differential equation, a first order Euler method is
chosen. Under this method, a component of the B-B force term can be
shown to be: 
\begin{equation}
\int_{0}^{\overline{t}}\frac{d\overline{v}}{d\overline{\tau }}\frac{d%
\overline{\tau }}{\sqrt{\left( \overline{t}-\overline{\tau }\right) }}%
=\sum\limits_{i=1}^{k-1}\frac{\overline{v}_{i+1}-\overline{v}_{i}}{\sqrt{%
\Delta \overline{t}}}\alpha \left( k,i\right) ,  \label{intediscre}
\end{equation}%
where we have assumed that $\Delta \overline{t}=\Delta \tau $ and defined $%
\alpha \left( k,i\right) =2\sqrt{k-i}-2\sqrt{k-1-i}$. 
After certain steps, one can prove that the overdamped M-R Eq. (\ref{overMR}%
), along the $z-$direction and in dimensional variables, acquires the
following discrete form for $k=3,...N$:%
\begin{eqnarray}
v_{z,k} &=&-\sqrt{\Delta t}\frac{c_{0}}{c_{1}}+\frac{c_{2}}{c_{1}}v_{z,k-1} 
\notag \\
&&-\frac{c_{3}}{c_{1}}\sum\limits_{i=1}^{k-2}\left( v_{z,i+1}-v_{z,i}\right)
\alpha \left( k,i\right) ,  \label{vzover}
\end{eqnarray}%
\begin{figure}
\includegraphics [width=7cm]{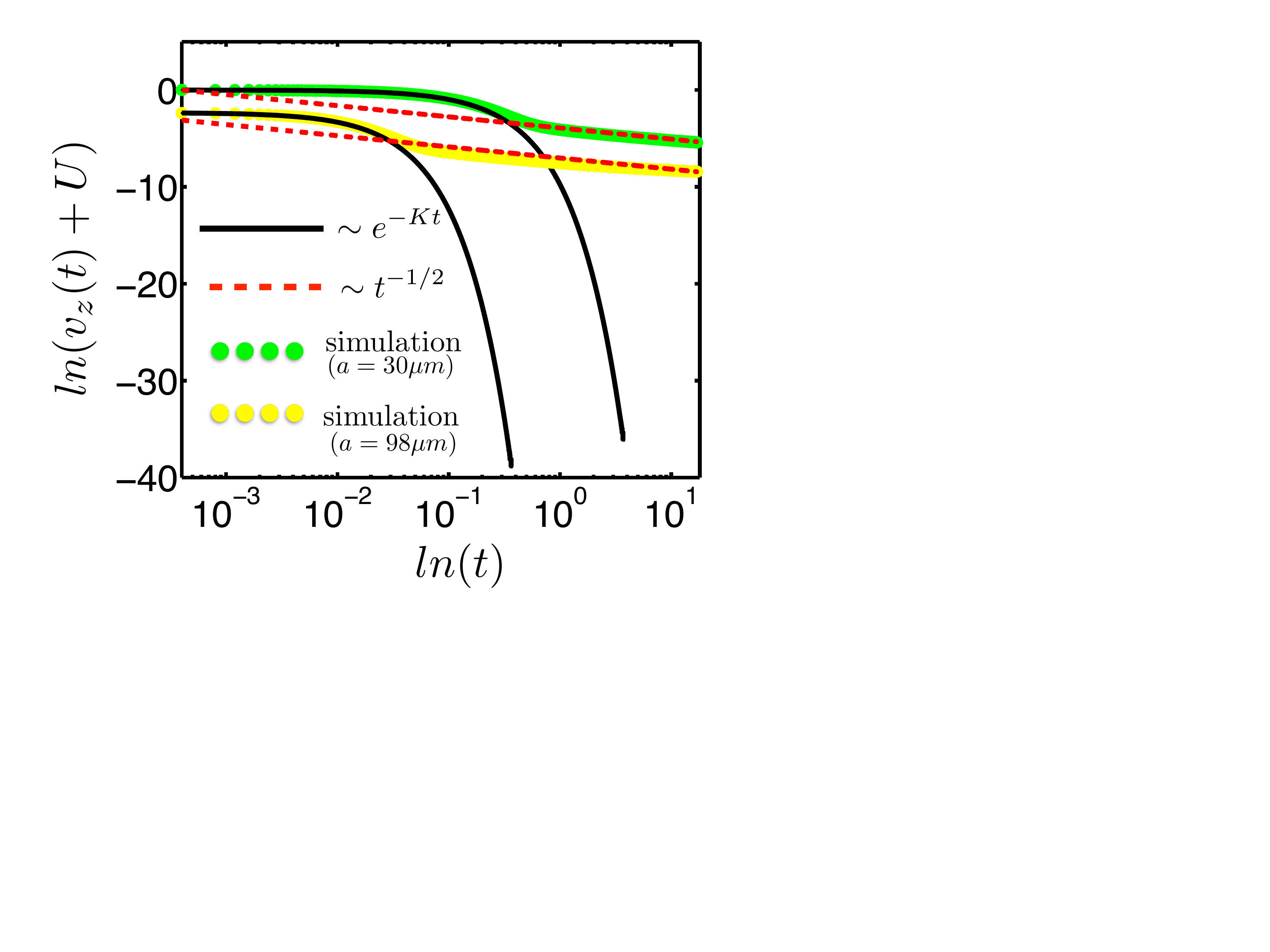}
\caption{Effect of the Boussinesq-Basset memory force term on the
sedimentation velocity ($v_{z}(0)=0$), for two spherical droplets reaching
their terminal velocity $U$. The black-solid lines indicate an exponential
decay towards $U$ when the B-B term is absent. The red-dashed lines indicate a $t^{-1/2}$
decay when the B-B term is present.}
\label{BM}
\end{figure}
where constants $c_{0},c_{1},c_{2},c_{3}$ are defined in Appendix \ref{A2}.
A similar expression as Eq. (\ref{vzover}) is obtained for the other spatial
components. In the case of larger droplets and from Table \ref{T1}, one
notices that the order-of-magnitude of all terms in Eq. (\ref{gen}) is the
same. Therefore, one has to solve for the full M-R Eq. (\ref{gen}). Using
the Euler method together with Eq. (\ref{intediscre}), one can show that the
discretized $z-$component of Eq. (\ref{gen}) in dimensional variables reads 
\begin{eqnarray}
v_{z,k} &=&-\frac{\Delta t}{b_{1}}Rg+\frac{b_{2}}{b_{1}}v_{z,k-1}  \notag \\
&&-\frac{b_{3}}{b_{1}}\sum\limits_{i=1}^{k-2}\left( v_{z,i+1}-v_{z,i}\right)
\alpha \left( k,i\right) ,  \label{VZFULL}
\end{eqnarray}%
where constants $R,b_{0},b_{1},b_{2},b_{3}$ are defined in Appendix \ref{A2}%
. The discretization of the other components in Eq. (\ref{gen}) is similar
to Eq. (\ref{VZFULL}). After posing Eq. (\ref{vzover}) and Eq.(\ref{VZFULL}%
), we are ready to find the droplets' dynamics under higher order terms.

Because of the external flows we have chosen and from the simulations in
Sec. \ref{no}, we observe that the dynamics mostly occurs along the $z-y$
plane, and that initial conditions are not relevant for long times (at least for $\mathbf{v}_{T}(0)=0$); thus from
now on, a 2D problem with initial conditions $v_{z}(0)=0$ and $%
v_{y}(0)=U_{max}$, will be considered. Equations (\ref{gen}) and (\ref%
{overMR}) are then numerically solved using the discretized scheme in Eq. (%
\ref{vzover}) and Eq. (\ref{VZFULL}), under the presence of a shear flow
(the other flows basically share the same features), and for two typical
exit initial speeds while talking \cite{AbkarianPNAS,1} namely, $1m/s$ and $%
5m/s$. The time-step used for solving Eq. (\ref{gen}) and Eq. (\ref{overMR}) was $4\times 10^{-4}s$
 and $5\times 10^{-3}s$, respectively. The results of
the simulations are shown in Figs. \ref{basset}(a)-(c). In those figures, $%
\Delta y=y_{MB}-y_{M}$, where $y_{MB}$ is the maximum travelled distance
along the $y$-direction by droplets of size $a\leq 20 \mu m$ and obtained
from Eq. (\ref{overMR}). For droplets of size $a> 20 \mu m$, $y_{MB}$
represents the maximum travelled distance along the $y$-direction obtained
after solving the full M-R equation (\ref{gen}); whereas $y_{M}$ represents
the solution directly obtained from Eqs. (\ref{xz}) and (\ref{y}). One can
observe that for a low wind speed, the effect of higher order terms barely
enhance the droplets' dispersion. However, for a wind speed of $5m/s$ and
for the smallest considered droplet, higher order terms can increase its dispersion
around 2 $m$. Figures \ref{basset}(a)-(c) also indicate that as the size of
the droplets increases, higher order terms effects become smaller until they finally 
disappear. 

The effect of first order terms and the full terms in the M-R equation, on
the flying time of spherical droplets as a function of the radius $a$ is also
analyzed. Using Eq. (\ref{xz}), the numerical solutions from Eq. (\ref%
{vzover}) and Eq. (\ref{VZFULL}), and defining $\Delta t_F=t_{FB}-t_{F}$;
where for $a\leq 20
\mu m$, $t_{FB}$ represents the flying time of a droplet  calculated using first order terms (Eq. (\ref{overMR})). For $a>
20 \mu m$, $t_{FB}$ represents the flying time  calculated using the full terms in the M-R
equation. On the other hand,  $t_{F}$ is the flying time directly obtained from Eqs. (\ref%
{xz}). This analysis is shown in Fig. \ref{distance}(d). It can be seen that 
$\Delta t_F$ increases as the droplets' sizes decrease. In fact, for $%
a=5\mu m$ there is a $0.8s$ flying time difference between the dynamics of
Eq. (\ref{gen}) and Eq. (\ref{SE}). This extra time also contributes to the
observed enhancement of dispersion of small droplets. A numerical analysis
on the convergence of $\Delta y=y_{MB}-y_{M}$, as the time-step $\Delta t$
in the simulations decreases, is also performed. Fig. \ref{basset}(f) shows
this convergence for a droplet of radius $30 \mu m$ after discretizing the
full M-R equation (\ref{gen}). As expected, a linear convergence can be
appreciated. The convergence for a droplet of radius $19 \mu m$ and after
using the overdamped M-R equation (\ref{overMR}) is shown in Fig. \ref%
{distance}(b). Once again a linear convergence is achieved. Therefore, the
latter analysis validates our employed first order numerical algorithm.

Finally, the implications of the Boussinesq-Basset memory force term on the
sedimentation velocity component $v_z(t)$ with $v_z(0)=0$ is also studied.
The numerical results are shown in Fig. \ref{BM}, which illustrates the
dynamics of $v_z(t)+U$ versus time in a log-log plot, and for two different spherical
droplets reaching their terminal velocity $U$. The black-solid lines belong
to an exponential decay of $v_z(t)$ towards $U$; whereas the red-dashed
lines indicate a $t^{-1/2}$ decay. It can be observed that for short times, $%
v_z(t)$ exponentially decays towards $U$; however, $v_z(t)$ decays 
according to the scaling $t^{-1/2}$ for long times. This is a rather
surprising result, since the order-of-magnitude of the B-B term is really
small. This $t^{-1/2}$ decay of the sedimenting velocity has been also
recently reported\cite{12}. Further consequences of the B-B term on the motion of
particles at low Reynolds numbers may be search for in future investigations.

\begin{figure}[tbp]
\includegraphics [width=8.5cm]{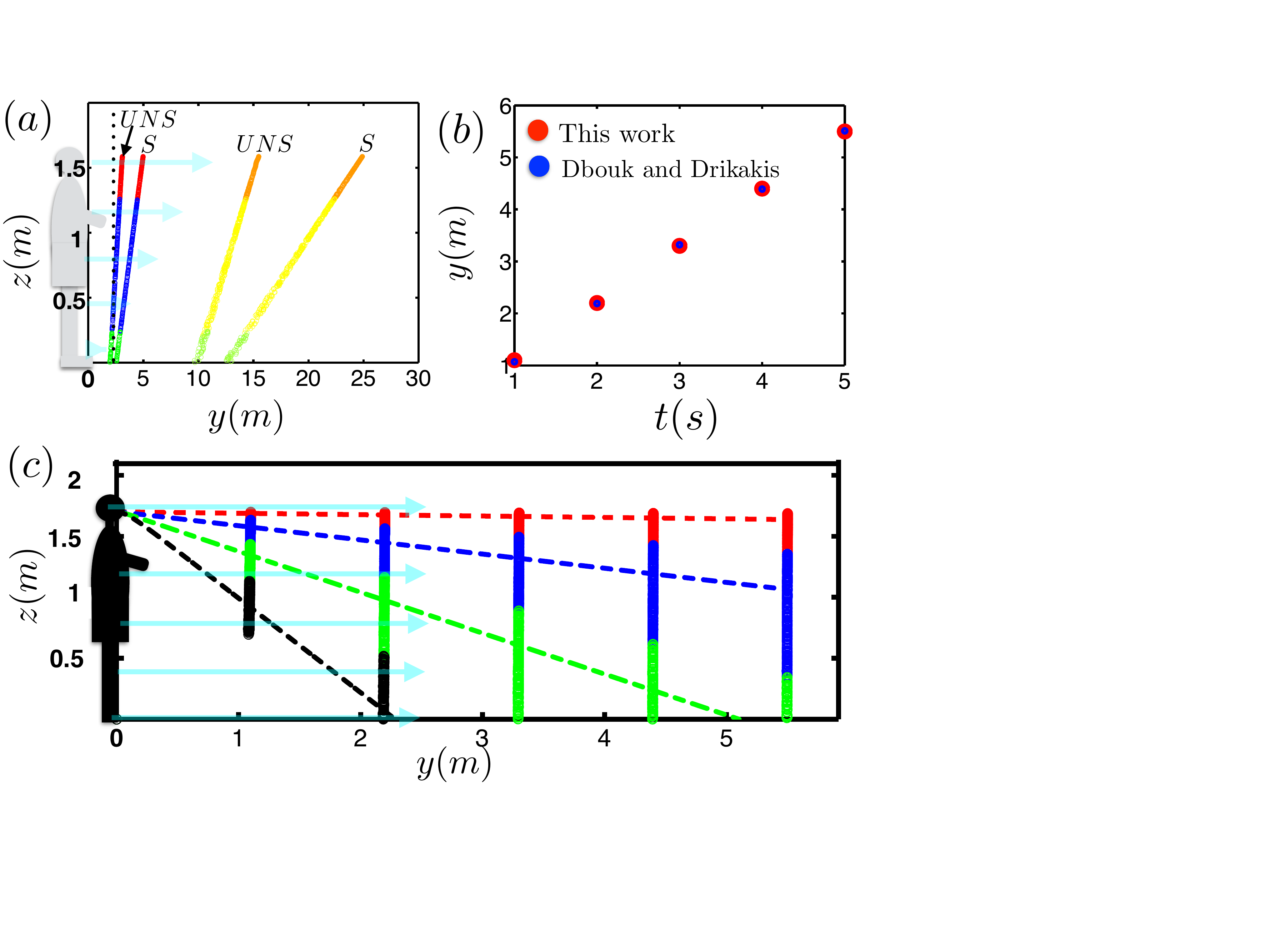}
\caption{(a) Droplets' distribution at $t=5s$ under a shear (S) and an
unsteady shear flow (UNS) at calm ($1m/s$) and moderate winds ($5m/s$).
The rest of the parameters are the same as in Fig. \protect\ref{model}.
Under these parameters, together with $RH=50\%$ and temperature les than
$30^oC$, evaporation does not play a role\protect\cite{2}. (b) Comparison of 
droplets' position under an uniform flow ($\{h,
U_{max},v_0\}=\{h=1.7m,1.1m/s,1m/s\}$) either using single particle dynamics
(this work), or using a more elaborated CFD analysis\protect\cite{2}. (c)
Droplets' distribution at $t=\{1s,2s,3s,4s,5s\}$, and linear paths followed
by some droplets of different sizes under an uniform flow. See Fig. \ref{model} for color code. }
\label{eva}
\end{figure}

\section{Discussions and conclusions}

\label{conclus}

According to Dbouk and Drikakis\cite{2} and others, evaporation and relative humidity
play a role on droplets' dispersion. Those factors may
reduce or increase the size of droplets and their cloud shape as it travels. Therefore, our results
could be improved to account for a rocket-like dynamics (drops varying
mass). Following Dbouk
and Drikakis\cite{2}, who considered a conjugated flow-heat-mass transfer
problem and CFD dynamics, it can be conclued that evaporation of droplets
takes place at high relative humidities, and high temperatures. Thus for the case
of countries with an annual average of relative humidity, $RH=50\%$, a wind speed of $4km/h=1.1m/s$%
, a temperature less than $30C^o$, and five seconds later since a cloud of
droplets originated, there will be a null evaporation\cite{2}. Other works also supporting a long time survival of infected droplets is Stadnytskyi {\it et al.}\cite{25}

Based
on this information, Fig. \ref{eva}(a) shows four droplets' distributions (cloud) five seconds later since the cloud originated  under a shear ($%
S$) or an unsteady shear ($UNS$) flow with $\{U_{max},v_0\}=\{1m/s,5m/s\}$.
The other  numerical parameters employed and the initial velocity distribution of  droplets are the
same as in Fig. \ref{model}. This figure indicates that droplets with $a>
75\mu m$ have already reached the floor, and that droplets of size $50\mu
m<a<75\mu m$ are about to reach the floor. However, the smallest droplets
under a calm wind ($1m/s$) for both $UNS$ and $S$, have surpassed the
social distancing (vertical black-dashed lines). The same droplets under a
moderate wind ($5m/s$) for both $UNS$ and $S$, reached $15m$ and $25m$,
 respectively. Thus from this figure, one can explicitly visualize a more real scale to
which people may be in risk of contagion. Fig. \ref{eva}(b) compares the
position of droplets' distribution (cloud) under an uniform flow ($\{h,
U_{max},v_0\}=\{h=1.7m,1.1m/s,1m/s\}$), either using single particle dynamics
(this work) or employing a more elaborated CFD analysis\cite{2}. Both methods result
in a similar cloud's position. Finally, using the latter parameters, the droplets'
distribution at $t=\{1s,2s,3s,4s,5s\}$ and some paths followed by droplets of different sizes are shown in Fig.\ref%
{eva}(c). These paths have a linear behavior since the cloud dynamics is
practically overdamped, implying $y(t)\approx U_{max}t$ and $z(t)\approx h-U t
$, thus particles will follow the function $z\approx h-(U/U_{max})y$.

In summary, using single particle dynamics, which has the advantage of requiring a
minimum computational cost, this paper provided an estimate
of the maximum dispersion of micrometric droplets generated after talking. Briefly, under conditions of a null evaporation and only five seconds later since droplets were originated, it was found that an unsteady shear  calm wind ($1m/s$) can disperse droplets beyond the social distancing, and until more than $15m$  when droplets are subject to a moderate unsteady shear flow ($5m/s$). As expected, an unsteady shear profile is less efficient to disperse droplets than a constant Poiseuille or shear profile. These constants profiles modeling a calm wind ($1m/s$) were found to provide a maximum dispersion   beyond $250m$ for the smallest droples ($5\mu m$). 
The effect of the Boussinesq--Basset force term
was also analyzed. Although its order-of magnitude is small, it was found to be enough
to change the behavior of the sedimentation velocity (from exponentially decaying towards its terminal velocity, to proportionally decaying as $t^{-1/2}$) of a micrometric
particle and slightly increase  droplets' dispersion and their flying time.

Future research would be to consider external flows in the presence of
buildings and to find the complex streamlines generated and their effects on
droplets' distribution. It may be inferred for example that the presence of
a corner on a common street, could generate stagnation points, that may risk
areas of infection, since those points could storage for a while infected
droplets. Furthermore, walls may induce three-dimensional flows that may
drag particles away from the floor, thus increasing its flying time and
hence their capability of traveling longer.
We hope this study helps people to be more aware about the effect of daily
wind currents on the propagation of potentially infected droplets. Based on
our findings of droplets easily dispersing beyond the social distancing when subject to wind currents, we recommend the use of masks\cite{linden} able to contain the
virus, as well as flex seal googles, since droplets dragged by wind may
reach eyes. We also recommend to wash all your wearing clothes and shoes,
and to shower after being out from home, since infected droplets may be
attached to clothes or hair\cite{30}. Direct exposure to wind currents,
mainly in crowded cities, should also be avoided. 

\section{Author contributios}

All authors contributed equally to this research.

\section{Acknowledgements}

M. S. thanks Consejo Nacional de Ciencia y Tecnologia, CONACyT for support.
M. S. dedicates this paper to the memory of his relatives Arturo Sandoval and Marcos Fernandez reached by this
pandemic.

\section{Data availability}

The data that support the findings of this study are available from the
corresponding author upon reasonable request.

\section{Appendix 1: Constants added}

\label{A1}

The constants $A,B,C,\gamma _{y},\gamma _{z},s_{x},s_{y},s_{z}$ appearing in
Eqs. (\ref{vx})-(\ref{y}) are defined as 
\begin{eqnarray}  \label{gamy}
A &=&\gamma _{z}\left( \frac{U_{\max }}{h}+U_{0}\left( 2s_{z}-h\right)
\right) ,  \label{A} \\
B &=&U_{0}s_{z}^{2}+\frac{\beta }{K}  \notag \\
&&+\left( U_{0}\left[ \frac{2g}{K^{2}}-h\right] +\frac{U_{\max }}{h}\right)
\left( \frac{g}{K^{2}}+s_{z}\right) ,  \label{B} \\
C &=&\frac{g}{K}\left[ U_{0}\left( 2\left( \frac{g}{K^{2}}+s_{z}\right)
-h\right) +\frac{U_{\max }}{h}\right] ,  \label{C} \\
\gamma _{y} &=&v_{y0}+U_{0}\left( \left( \frac{\gamma _{z}}{K}\right)
^{2}-s_{z}^{2}\right)   \notag \\
&&-\left( U_{0}\left( \frac{2g}{K^{2}}-h\right) +\frac{U_{\max }}{h}\right)
\left( \frac{g}{K^{2}}+s_{z}\right) -\frac{\beta }{K},  \notag \\
&& \\
\gamma _{z} &=&v_{z0}+\frac{g}{K},s_{x}=\frac{v_{x0}}{K},s_{z}=h+\frac{%
\gamma _{z}}{K},  \label{gamz} \\
s_{y} &=&-\frac{1}{K}\left[ \frac{A}{K}-\gamma _{y}+\frac{U_{0}\gamma _{z}}{%
K^{2}}\left( \frac{\gamma _{z}}{2}-\frac{2g}{K}\right) \right] .  \label{sy}
\end{eqnarray}

\section{Appendix 2: Numerical part for the B-B equation}

\label{A2} The constants $c_{0},c_{1},c_{2},c_{3}$ appearing in Eq. (\ref%
{vzover}) are defined as 
\begin{eqnarray}
c_{0} &=&\frac{g\left( m-m_{a}\right) }{R_{T}},  \label{co} \\
c_{1} &=&\frac{2a}{\sqrt{\pi \upsilon _{a}}}+\sqrt{\Delta t},  \label{c1} \\
c_{2} &=&\frac{2a}{\sqrt{\pi \upsilon _{a}}},c_{3}=\frac{c_{2}}{2},
\label{c23}
\end{eqnarray}%
whereas constants $R,b_{1},b_{2},b_{3}$ appearing in Eq. (\ref{VZFULL}) are 
\begin{eqnarray}
R &=&(1-\lambda )/D\text{ with }D=1+\lambda /2,  \label{D} \\
b_{1} &=&\left( 1+\sqrt{\Delta t}\frac{2Ka}{D\sqrt{\pi \upsilon _{a}}}%
\right) ,  \label{b1} \\
b_{2} &=&\sqrt{\Delta t}\frac{2Ka}{D\sqrt{\pi \upsilon _{a}}}-\Delta t\frac{K%
}{D}+1,  \label{b2} \\
b_{3} &=&\sqrt{\Delta t}\frac{Ka}{D\sqrt{\pi \upsilon _{a}}}.  \label{b3}
\end{eqnarray}

\bibliography{covidref}

\providecommand{\noopsort}[1]{}\providecommand{\singleletter}[1]{#1}%
\begin{thebibliography}{28}%
\makeatletter
\providecommand \@ifxundefined [1]{%
 \@ifx{#1\undefined}
}%
\providecommand \@ifnum [1]{%
 \ifnum #1\expandafter \@firstoftwo
 \else \expandafter \@secondoftwo
 \fi
}%
\providecommand \@ifx [1]{%
 \ifx #1\expandafter \@firstoftwo
 \else \expandafter \@secondoftwo
 \fi
}%
\providecommand \natexlab [1]{#1}%
\providecommand \enquote  [1]{``#1''}%
\providecommand \bibnamefont  [1]{#1}%
\providecommand \bibfnamefont [1]{#1}%
\providecommand \citenamefont [1]{#1}%
\providecommand \href@noop [0]{\@secondoftwo}%
\providecommand \href [0]{\begingroup \@sanitize@url \@href}%
\providecommand \@href[1]{\@@startlink{#1}\@@href}%
\providecommand \@@href[1]{\endgroup#1\@@endlink}%
\providecommand \@sanitize@url [0]{\catcode `\\12\catcode `\$12\catcode
  `\&12\catcode `\#12\catcode `\^12\catcode `\_12\catcode `\%12\relax}%
\providecommand \@@startlink[1]{}%
\providecommand \@@endlink[0]{}%
\providecommand \url  [0]{\begingroup\@sanitize@url \@url }%
\providecommand \@url [1]{\endgroup\@href {#1}{\urlprefix }}%
\providecommand \urlprefix  [0]{URL }%
\providecommand \Eprint [0]{\href }%
\providecommand \doibase [0]{http://dx.doi.org/}%
\providecommand \selectlanguage [0]{\@gobble}%
\providecommand \bibinfo  [0]{\@secondoftwo}%
\providecommand \bibfield  [0]{\@secondoftwo}%
\providecommand \translation [1]{[#1]}%
\providecommand \BibitemOpen [0]{}%
\providecommand \bibitemStop [0]{}%
\providecommand \bibitemNoStop [0]{.\EOS\space}%
\providecommand \EOS [0]{\spacefactor3000\relax}%
\providecommand \BibitemShut  [1]{\csname bibitem#1\endcsname}%
\let\auto@bib@innerbib\@empty
\bibitem [{\citenamefont {Zhang}\ \emph {et~al.}(2020)\citenamefont {Zhang},
  \citenamefont {Li}, \citenamefont {Zhang}, \citenamefont {Wang},\ and\
  \citenamefont {Molina}}]{molina}%
  \BibitemOpen
  \bibfield  {author} {\bibinfo {author} {\bibfnamefont {R.}~\bibnamefont
  {Zhang}}, \bibinfo {author} {\bibfnamefont {Y.}~\bibnamefont {Li}}, \bibinfo
  {author} {\bibfnamefont {A.~L.}\ \bibnamefont {Zhang}}, \bibinfo {author}
  {\bibfnamefont {Y.}~\bibnamefont {Wang}}, \ and\ \bibinfo {author}
  {\bibfnamefont {M.~J.}\ \bibnamefont {Molina}},\ }\bibfield  {title}
  {\enquote {\bibinfo {title} {Identifying airborne transmission as the
  dominant route for the spread of covid-19},}\ }\href {\doibase
  10.1073/pnas.2009637117} {\bibfield  {journal} {\bibinfo  {journal}
  {Proceedings of the National Academy of Sciences}\ }\textbf {\bibinfo
  {volume} {117}},\ \bibinfo {pages} {14857--14863} (\bibinfo {year} {2020})},\
  \Eprint
  {http://arxiv.org/abs/https://www.pnas.org/content/117/26/14857.full.pdf}
  {https://www.pnas.org/content/117/26/14857.full.pdf} \BibitemShut {NoStop}%
\bibitem [{\citenamefont {Mittal}, \citenamefont {Ni},\ and\ \citenamefont
  {Seo}(2020)}]{6}%
  \BibitemOpen
  \bibfield  {author} {\bibinfo {author} {\bibfnamefont {R.}~\bibnamefont
  {Mittal}}, \bibinfo {author} {\bibfnamefont {R.}~\bibnamefont {Ni}}, \ and\
  \bibinfo {author} {\bibfnamefont {J.-H.}\ \bibnamefont {Seo}},\ }\bibfield
  {title} {\enquote {\bibinfo {title} {The flow physics of covid-19},}\
  }\href@noop {} {\bibfield  {journal} {\bibinfo  {journal} {Journal of Fluid
  Mechanics}\ }\textbf {\bibinfo {volume} {{\bf 894}}},\ \bibinfo {pages} {F2}
  (\bibinfo {year} {2020})}\BibitemShut {NoStop}%
\bibitem [{\citenamefont {Abkarian}\ and\ \citenamefont {Stone}(2020)}]{11}%
  \BibitemOpen
  \bibfield  {author} {\bibinfo {author} {\bibfnamefont {M.}~\bibnamefont
  {Abkarian}}\ and\ \bibinfo {author} {\bibfnamefont {H.~A.}\ \bibnamefont
  {Stone}},\ }\bibfield  {title} {\enquote {\bibinfo {title} {Stretching and
  break-up of saliva filaments during speech: A route for pathogen
  aerosolization and its potential mitigation},}\ }\href@noop {} {\bibfield
  {journal} {\bibinfo  {journal} {Physical Review Fluids}\ }\textbf {\bibinfo
  {volume} {{\bf 5}}},\ \bibinfo {pages} {102301} (\bibinfo {year}
  {2020})}\BibitemShut {NoStop}%
\bibitem [{\citenamefont {Abkarian}\ \emph {et~al.}(2020)\citenamefont
  {Abkarian}, \citenamefont {Mendez}, \citenamefont {Xue}, \citenamefont
  {Yang},\ and\ \citenamefont {Stone}}]{AbkarianPNAS}%
  \BibitemOpen
  \bibfield  {author} {\bibinfo {author} {\bibfnamefont {M.}~\bibnamefont
  {Abkarian}}, \bibinfo {author} {\bibfnamefont {S.}~\bibnamefont {Mendez}},
  \bibinfo {author} {\bibfnamefont {N.}~\bibnamefont {Xue}}, \bibinfo {author}
  {\bibfnamefont {F.}~\bibnamefont {Yang}}, \ and\ \bibinfo {author}
  {\bibfnamefont {H.~A.}\ \bibnamefont {Stone}},\ }\bibfield  {title} {\enquote
  {\bibinfo {title} {Speech can produce jet-like transport relevant to
  asymptomatic spreading of virus},}\ }\href {\doibase 10.1073/pnas.2012156117}
  {\bibfield  {journal} {\bibinfo  {journal} {Proceedings of the National
  Academy of Sciences}\ }\textbf {\bibinfo {volume} {117}},\ \bibinfo {pages}
  {25237--25245} (\bibinfo {year} {2020})},\ \Eprint
  {http://arxiv.org/abs/https://www.pnas.org/content/117/41/25237.full.pdf}
  {https://www.pnas.org/content/117/41/25237.full.pdf} \BibitemShut {NoStop}%
\bibitem [{\citenamefont {Wright}(2020)}]{Katy}%
  \BibitemOpen
  \bibfield  {author} {\bibinfo {author} {\bibfnamefont {K.}~\bibnamefont
  {Wright}},\ }\bibfield  {title} {\enquote {\bibinfo {title} {How talking
  spreads viruses},}\ }\href@noop {} {\bibfield  {journal} {\bibinfo  {journal}
  {Physics}\ }\textbf {\bibinfo {volume} {{\bf 13}}},\ \bibinfo {pages} {195}
  (\bibinfo {year} {2020})}\BibitemShut {NoStop}%
\bibitem [{\citenamefont {Bourouiba}(2020)}]{Bourouiba}%
  \BibitemOpen
  \bibfield  {author} {\bibinfo {author} {\bibfnamefont {L.}~\bibnamefont
  {Bourouiba}},\ }\bibfield  {title} {\enquote {\bibinfo {title} {{Turbulent
  Gas Clouds and Respiratory Pathogen Emissions: Potential Implications for
  Reducing Transmission of COVID-19}},}\ }\href {\doibase
  10.1001/jama.2020.4756} {\bibfield  {journal} {\bibinfo  {journal} {JAMA}\
  }\textbf {\bibinfo {volume} {323}},\ \bibinfo {pages} {1837--1838} (\bibinfo
  {year} {2020})}\BibitemShut {NoStop}%
\bibitem [{\citenamefont {Jones}\ \emph {et~al.}(2020)\citenamefont {Jones},
  \citenamefont {Qureshi}, \citenamefont {Temple}, \citenamefont {Larwood},
  \citenamefont {Greenhalgh},\ and\ \citenamefont {Bourouiba}}]{Jones}%
  \BibitemOpen
  \bibfield  {author} {\bibinfo {author} {\bibfnamefont {N.~R.}\ \bibnamefont
  {Jones}}, \bibinfo {author} {\bibfnamefont {Z.~U.}\ \bibnamefont {Qureshi}},
  \bibinfo {author} {\bibfnamefont {R.~J.}\ \bibnamefont {Temple}}, \bibinfo
  {author} {\bibfnamefont {J.~P.~J.}\ \bibnamefont {Larwood}}, \bibinfo
  {author} {\bibfnamefont {T.}~\bibnamefont {Greenhalgh}}, \ and\ \bibinfo
  {author} {\bibfnamefont {L.}~\bibnamefont {Bourouiba}},\ }\bibfield  {title}
  {\enquote {\bibinfo {title} {Two metres or one: what is the evidence for
  physical distancing in covid-19?}}\ }\href {\doibase 10.1136/bmj.m3223}
  {\bibfield  {journal} {\bibinfo  {journal} {BMJ}\ }\textbf {\bibinfo {volume}
  {370}} (\bibinfo {year} {2020}),\ 10.1136/bmj.m3223},\ \Eprint
  {http://arxiv.org/abs/https://www.bmj.com/content/370/bmj.m3223.full.pdf}
  {https://www.bmj.com/content/370/bmj.m3223.full.pdf} \BibitemShut {NoStop}%
\bibitem [{\citenamefont {Bhagat}\ \emph {et~al.}(2020)\citenamefont {Bhagat},
  \citenamefont {Davies~Wykes}, \citenamefont {Dalziel},\ and\ \citenamefont
  {Linden}}]{linden}%
  \BibitemOpen
  \bibfield  {author} {\bibinfo {author} {\bibfnamefont {R.~K.}\ \bibnamefont
  {Bhagat}}, \bibinfo {author} {\bibfnamefont {M.~S.}\ \bibnamefont
  {Davies~Wykes}}, \bibinfo {author} {\bibfnamefont {S.~B.}\ \bibnamefont
  {Dalziel}}, \ and\ \bibinfo {author} {\bibfnamefont {P.~F.}\ \bibnamefont
  {Linden}},\ }\bibfield  {title} {\enquote {\bibinfo {title} {Effects of
  ventilation on the indoor spread of covid-19},}\ }\href {\doibase
  10.1017/jfm.2020.720} {\bibfield  {journal} {\bibinfo  {journal} {Journal of
  Fluid Mechanics}\ }\textbf {\bibinfo {volume} {903}},\ \bibinfo {pages} {F1}
  (\bibinfo {year} {2020})}\BibitemShut {NoStop}%
\bibitem [{\citenamefont {Park}\ \emph {et~al.}(2020)\citenamefont {Park},
  \citenamefont {Kim}, \citenamefont {Yi}, \citenamefont {Lee}, \citenamefont
  {Na}, \citenamefont {Kim},\ and\ \citenamefont {Jeong}}]{park}%
  \BibitemOpen
  \bibfield  {author} {\bibinfo {author} {\bibfnamefont {S.}~\bibnamefont
  {Park}}, \bibinfo {author} {\bibfnamefont {Y.}~\bibnamefont {Kim}}, \bibinfo
  {author} {\bibfnamefont {S.}~\bibnamefont {Yi}}, \bibinfo {author}
  {\bibfnamefont {S.}~\bibnamefont {Lee}}, \bibinfo {author} {\bibfnamefont
  {B.}~\bibnamefont {Na}}, \bibinfo {author} {\bibfnamefont {C.}~\bibnamefont
  {Kim}}, \ and\ \bibinfo {author} {\bibfnamefont {E.}~\bibnamefont {Jeong}},\
  }\bibfield  {title} {\enquote {\bibinfo {title} {Coronavirus disease outbreak
  in call center, south korea},}\ }\href@noop {} {\bibfield  {journal}
  {\bibinfo  {journal} {Emerging Infectious Diseases}\ }\textbf {\bibinfo
  {volume} {26(8)}},\ \bibinfo {pages} {1666--1670} (\bibinfo {year}
  {(2020)})}\BibitemShut {NoStop}%
\bibitem [{\citenamefont {Cummins}\ \emph {et~al.}(2020)\citenamefont
  {Cummins}, \citenamefont {Ajayi}, \citenamefont {Mehendale}, \citenamefont
  {Gabl},\ and\ \citenamefont {Viola}}]{1}%
  \BibitemOpen
  \bibfield  {author} {\bibinfo {author} {\bibfnamefont {C.~P.}\ \bibnamefont
  {Cummins}}, \bibinfo {author} {\bibfnamefont {O.~J.}\ \bibnamefont {Ajayi}},
  \bibinfo {author} {\bibfnamefont {F.~V.}\ \bibnamefont {Mehendale}}, \bibinfo
  {author} {\bibfnamefont {R.}~\bibnamefont {Gabl}}, \ and\ \bibinfo {author}
  {\bibfnamefont {I.~M.}\ \bibnamefont {Viola}},\ }\bibfield  {title} {\enquote
  {\bibinfo {title} {The dispersion of spherical droplets in source-sink flows
  and their relevance to the covid-19 pandemic},}\ }\href@noop {} {\bibfield
  {journal} {\bibinfo  {journal} {Physics of Fluids}\ }\textbf {\bibinfo
  {volume} {{\bf 32}}},\ \bibinfo {pages} {083302} (\bibinfo {year}
  {2020})}\BibitemShut {NoStop}%
\bibitem [{\citenamefont {Dbouk}\ and\ \citenamefont {Drikakis}(2020)}]{2}%
  \BibitemOpen
  \bibfield  {author} {\bibinfo {author} {\bibfnamefont {T.}~\bibnamefont
  {Dbouk}}\ and\ \bibinfo {author} {\bibfnamefont {D.}~\bibnamefont
  {Drikakis}},\ }\bibfield  {title} {\enquote {\bibinfo {title} {Weather impact
  on airborne coronavirus survival},}\ }\href@noop {} {\bibfield  {journal}
  {\bibinfo  {journal} {Physics of Fluids}\ }\textbf {\bibinfo {volume} {{\bf
  32}}},\ \bibinfo {pages} {093312} (\bibinfo {year} {2020})}\BibitemShut
  {NoStop}%
\bibitem [{\citenamefont {Blocken~B}()}]{28}%
  \BibitemOpen
  \bibfield  {author} {\bibinfo {author} {\bibfnamefont {v.~D. T. M.~T.}\
  \bibnamefont {Blocken~B}, \bibfnamefont {Malizia~F}},\ }\bibfield  {title}
  {\enquote {\bibinfo {title} {Towards aerodynamically equivalent covid19
  $1.5m$ social distancing for walking and running. 2020},}\ }\href@noop {}
  {\bibinfo  {journal} {Pre-print Available online: {
  http://www.urbanphysics.net/COVID19$\_$Aero$\_$Paper.pdf}. Accessed 26 Jan
  2021}\ }\BibitemShut {NoStop}%
\bibitem [{\citenamefont {Dhand}\ and\ \citenamefont {Li}(2020)}]{26}%
  \BibitemOpen
\bibfield  {journal} {  }\bibfield  {author} {\bibinfo {author} {\bibfnamefont
  {R.}~\bibnamefont {Dhand}}\ and\ \bibinfo {author} {\bibfnamefont
  {J.}~\bibnamefont {Li}},\ }\bibfield  {title} {\enquote {\bibinfo {title}
  {Coughs and sneezes: their role in transmission of respiratory viral
  infections, including sars-cov-2},}\ }\href@noop {} {\bibfield  {journal}
  {\bibinfo  {journal} {American journal of respiratory and critical care
  medicine}\ }\textbf {\bibinfo {volume} {{\bf 202}}},\ \bibinfo {pages}
  {651--659} (\bibinfo {year} {2020})}\BibitemShut {NoStop}%
\bibitem [{\citenamefont {Feng}\ \emph {et~al.}(2020)\citenamefont {Feng},
  \citenamefont {Marchal}, \citenamefont {Sperry},\ and\ \citenamefont
  {Yi}}]{30}%
  \BibitemOpen
  \bibfield  {author} {\bibinfo {author} {\bibfnamefont {Y.}~\bibnamefont
  {Feng}}, \bibinfo {author} {\bibfnamefont {T.}~\bibnamefont {Marchal}},
  \bibinfo {author} {\bibfnamefont {T.}~\bibnamefont {Sperry}}, \ and\ \bibinfo
  {author} {\bibfnamefont {H.}~\bibnamefont {Yi}},\ }\bibfield  {title}
  {\enquote {\bibinfo {title} {Influence of wind and relative humidity on the
  social distancing effectiveness to prevent covid-19 airborne transmission: A
  numerical study},}\ }\href@noop {} {\bibfield  {journal} {\bibinfo  {journal}
  {Journal of aerosol science}\ }\textbf {\bibinfo {volume} {{\bf }}},\
  \bibinfo {pages} {105585} (\bibinfo {year} {2020})}\BibitemShut {NoStop}%
\bibitem [{\citenamefont {Maxey}\ and\ \citenamefont {Riley}(1983)}]{7}%
  \BibitemOpen
  \bibfield  {author} {\bibinfo {author} {\bibfnamefont {M.~R.}\ \bibnamefont
  {Maxey}}\ and\ \bibinfo {author} {\bibfnamefont {J.~J.}\ \bibnamefont
  {Riley}},\ }\bibfield  {title} {\enquote {\bibinfo {title} {Equation of
  motion for a small rigid sphere in a nonuniform flow},}\ }\href@noop {}
  {\bibfield  {journal} {\bibinfo  {journal} {The Physics of Fluids}\ }\textbf
  {\bibinfo {volume} {{\bf 26}}},\ \bibinfo {pages} {883--889} (\bibinfo {year}
  {1983})}\BibitemShut {NoStop}%
\bibitem [{\citenamefont {Fontes}\ \emph {et~al.}(2020)\citenamefont {Fontes},
  \citenamefont {Reyes}, \citenamefont {Ahmed},\ and\ \citenamefont
  {Kinzel}}]{3}%
  \BibitemOpen
  \bibfield  {author} {\bibinfo {author} {\bibfnamefont {D.}~\bibnamefont
  {Fontes}}, \bibinfo {author} {\bibfnamefont {J.}~\bibnamefont {Reyes}},
  \bibinfo {author} {\bibfnamefont {K.}~\bibnamefont {Ahmed}}, \ and\ \bibinfo
  {author} {\bibfnamefont {M.}~\bibnamefont {Kinzel}},\ }\bibfield  {title}
  {\enquote {\bibinfo {title} {A study of fluid dynamics and human physiology
  factors driving droplet dispersion from a human sneeze},}\ }\href@noop {}
  {\bibfield  {journal} {\bibinfo  {journal} {Physics of Fluids}\ }\textbf
  {\bibinfo {volume} {{\bf 32}}},\ \bibinfo {pages} {111904} (\bibinfo {year}
  {2020})}\BibitemShut {NoStop}%
\bibitem [{\citenamefont {Prasath}, \citenamefont {Vasan},\ and\ \citenamefont
  {Govindarajan}(2019)}]{12}%
  \BibitemOpen
  \bibfield  {author} {\bibinfo {author} {\bibfnamefont {S.~G.}\ \bibnamefont
  {Prasath}}, \bibinfo {author} {\bibfnamefont {V.}~\bibnamefont {Vasan}}, \
  and\ \bibinfo {author} {\bibfnamefont {R.}~\bibnamefont {Govindarajan}},\
  }\bibfield  {title} {\enquote {\bibinfo {title} {{Accurate solution method
  for the Maxey-Riley equation, and the effects of Basset history}},}\
  }\href@noop {} {\bibfield  {journal} {\bibinfo  {journal} {Journal of Fluid
  Mechanics}\ }\textbf {\bibinfo {volume} {{\bf 868}}},\ \bibinfo {pages}
  {428--460} (\bibinfo {year} {2019})}\BibitemShut {NoStop}%
\bibitem [{\citenamefont {Langlois}, \citenamefont {Farazmand},\ and\
  \citenamefont {Haller}(2015)}]{15}%
  \BibitemOpen
  \bibfield  {author} {\bibinfo {author} {\bibfnamefont {G.~P.}\ \bibnamefont
  {Langlois}}, \bibinfo {author} {\bibfnamefont {M.}~\bibnamefont {Farazmand}},
  \ and\ \bibinfo {author} {\bibfnamefont {G.}~\bibnamefont {Haller}},\
  }\bibfield  {title} {\enquote {\bibinfo {title} {Asymptotic dynamics of
  inertial particles with memory},}\ }\href@noop {} {\bibfield  {journal}
  {\bibinfo  {journal} {Journal of nonlinear science}\ }\textbf {\bibinfo
  {volume} {{\bf 25}}},\ \bibinfo {pages} {1225--1255} (\bibinfo {year}
  {2015})}\BibitemShut {NoStop}%
\bibitem [{\citenamefont {Guseva}\ and\ \citenamefont {andand
  Tam\'as~T\'el}(2013)}]{20}%
  \BibitemOpen
  \bibfield  {author} {\bibinfo {author} {\bibfnamefont {K.}~\bibnamefont
  {Guseva}}\ and\ \bibinfo {author} {\bibfnamefont {U.~F.}\ \bibnamefont
  {andand Tam\'as~T\'el}},\ }\bibfield  {title} {\enquote {\bibinfo {title}
  {Influence of the history force on inertial particle advection: Gravitational
  effects and horizontal diffusion},}\ }\href@noop {} {\bibfield  {journal}
  {\bibinfo  {journal} {Physical Review E}\ }\textbf {\bibinfo {volume} {{\bf
  88}}},\ \bibinfo {pages} {042909} (\bibinfo {year} {2013})}\BibitemShut
  {NoStop}%
\bibitem [{\citenamefont {Haller}(2019)}]{22}%
  \BibitemOpen
  \bibfield  {author} {\bibinfo {author} {\bibfnamefont {G.}~\bibnamefont
  {Haller}},\ }\bibfield  {title} {\enquote {\bibinfo {title} {Solving the
  inertial particle equation with memory},}\ }\href@noop {} {\bibfield
  {journal} {\bibinfo  {journal} {Journal of Fluid Mechanics}\ }\textbf
  {\bibinfo {volume} {{\bf 874}}},\ \bibinfo {pages} {1--4} (\bibinfo {year}
  {2019})}\BibitemShut {NoStop}%
\bibitem [{\citenamefont {Tatom}(1988)}]{23}%
  \BibitemOpen
  \bibfield  {author} {\bibinfo {author} {\bibfnamefont {F.}~\bibnamefont
  {Tatom}},\ }\bibfield  {title} {\enquote {\bibinfo {title} {The basset term
  as a semiderivative},}\ }\href@noop {} {\bibfield  {journal} {\bibinfo
  {journal} {Applied Scientific Research}\ }\textbf {\bibinfo {volume} {{\bf
  45}}},\ \bibinfo {pages} {283--285} (\bibinfo {year} {1988})}\BibitemShut
  {NoStop}%
\bibitem [{\citenamefont {van Hinsberg}, \citenamefont {ten Thije~Boonkkamp},\
  and\ \citenamefont {Clercx}(2011)}]{14}%
  \BibitemOpen
  \bibfield  {author} {\bibinfo {author} {\bibfnamefont {M.}~\bibnamefont {van
  Hinsberg}}, \bibinfo {author} {\bibfnamefont {J.}~\bibnamefont {ten
  Thije~Boonkkamp}}, \ and\ \bibinfo {author} {\bibfnamefont {H.}~\bibnamefont
  {Clercx}},\ }\bibfield  {title} {\enquote {\bibinfo {title} {An efficient,
  second order method for the approximation of the basset history force},}\
  }\href@noop {} {\bibfield  {journal} {\bibinfo  {journal} {Journal of
  Computational Physics}\ }\textbf {\bibinfo {volume} {{\bf 230}}},\ \bibinfo
  {pages} {1465 -- 1478} (\bibinfo {year} {2011})}\BibitemShut {NoStop}%
\bibitem [{\citenamefont {Alexander}(2004)}]{17}%
  \BibitemOpen
  \bibfield  {author} {\bibinfo {author} {\bibfnamefont {P.}~\bibnamefont
  {Alexander}},\ }\bibfield  {title} {\enquote {\bibinfo {title} {High order
  computation of the history term in the equation of motion for a spherical
  particle in a fluid},}\ }\href@noop {} {\bibfield  {journal} {\bibinfo
  {journal} {Journal of Scientific Computing}\ }\textbf {\bibinfo {volume}
  {{\bf 21}}},\ \bibinfo {pages} {129--143} (\bibinfo {year}
  {2004})}\BibitemShut {NoStop}%
\bibitem [{\citenamefont {Daitche}\ and\ \citenamefont {T\'el}(2014)}]{5}%
  \BibitemOpen
  \bibfield  {author} {\bibinfo {author} {\bibfnamefont {A.}~\bibnamefont
  {Daitche}}\ and\ \bibinfo {author} {\bibfnamefont {T.}~\bibnamefont
  {T\'el}},\ }\bibfield  {title} {\enquote {\bibinfo {title} {Memory effects in
  chaotic advection of inertial particles},}\ }\href@noop {} {\bibfield
  {journal} {\bibinfo  {journal} {New Journal of Physics}\ }\textbf {\bibinfo
  {volume} {{\bf 16}}},\ \bibinfo {pages} {073008} (\bibinfo {year}
  {2014})}\BibitemShut {NoStop}%
\bibitem [{\citenamefont {Daitche}\ and\ \citenamefont {T{\'e}l}(2011)}]{21}%
  \BibitemOpen
  \bibfield  {author} {\bibinfo {author} {\bibfnamefont {A.}~\bibnamefont
  {Daitche}}\ and\ \bibinfo {author} {\bibfnamefont {T.}~\bibnamefont
  {T{\'e}l}},\ }\bibfield  {title} {\enquote {\bibinfo {title} {Memory effects
  are relevant for chaotic advection of inertial particles},}\ }\href@noop {}
  {\bibfield  {journal} {\bibinfo  {journal} {Physical review letters}\
  }\textbf {\bibinfo {volume} {{\bf 107}}},\ \bibinfo {pages} {244501}
  (\bibinfo {year} {2011})}\BibitemShut {NoStop}%
\bibitem [{\citenamefont {Candelier}, \citenamefont {Angilella},\ and\
  \citenamefont {Souhar}(2004)}]{13}%
  \BibitemOpen
  \bibfield  {author} {\bibinfo {author} {\bibfnamefont {F.}~\bibnamefont
  {Candelier}}, \bibinfo {author} {\bibfnamefont {J.~R.}\ \bibnamefont
  {Angilella}}, \ and\ \bibinfo {author} {\bibfnamefont {M.}~\bibnamefont
  {Souhar}},\ }\bibfield  {title} {\enquote {\bibinfo {title} {On the effect of
  the boussinesq-basset force on the radial migration of a stokes particle in a
  vortex},}\ }\href@noop {} {\bibfield  {journal} {\bibinfo  {journal} {Physics
  of Fluids}\ }\textbf {\bibinfo {volume} {{\bf 16}}},\ \bibinfo {pages}
  {1765--1776} (\bibinfo {year} {2004})}\BibitemShut {NoStop}%
\bibitem [{\citenamefont {Mordant}\ and\ \citenamefont {Pinton}(2000)}]{27}%
  \BibitemOpen
  \bibfield  {author} {\bibinfo {author} {\bibfnamefont {N.}~\bibnamefont
  {Mordant}}\ and\ \bibinfo {author} {\bibfnamefont {J.-F.}\ \bibnamefont
  {Pinton}},\ }\bibfield  {title} {\enquote {\bibinfo {title} {Velocity
  measurement of a settling sphere},}\ }\href@noop {} {\bibfield  {journal}
  {\bibinfo  {journal} {The European Physical Journal B-Condensed Matter and
  Complex Systems}\ }\textbf {\bibinfo {volume} {{\bf 18}}},\ \bibinfo {pages}
  {343--352} (\bibinfo {year} {2000})}\BibitemShut {NoStop}%
\bibitem [{\citenamefont {Stadnytskyi}\ \emph {et~al.}(2020)\citenamefont
  {Stadnytskyi}, \citenamefont {Bax}, \citenamefont {Bax},\ and\ \citenamefont
  {Anfinrud}}]{25}%
  \BibitemOpen
  \bibfield  {author} {\bibinfo {author} {\bibfnamefont {V.}~\bibnamefont
  {Stadnytskyi}}, \bibinfo {author} {\bibfnamefont {C.~E.}\ \bibnamefont
  {Bax}}, \bibinfo {author} {\bibfnamefont {A.}~\bibnamefont {Bax}}, \ and\
  \bibinfo {author} {\bibfnamefont {P.}~\bibnamefont {Anfinrud}},\ }\bibfield
  {title} {\enquote {\bibinfo {title} {The airborne lifetime of small speech
  droplets and their potential importance in sars-cov-2 transmission},}\
  }\href@noop {} {\bibfield  {journal} {\bibinfo  {journal} {Proceedings of the
  National Academy of Sciences}\ }\textbf {\bibinfo {volume} {{\bf 117}}},\
  \bibinfo {pages} {11875--11877} (\bibinfo {year} {2020})}\BibitemShut
  {NoStop}%
\end{thebibliography}%

\end{document}